# Big Data in Astroinformatics -
# Compression of Scanned Astronomical Photographic Plates

Vasil Kolev

IICT, BAS,

Sofia, Bulgaria

e-mail: kolev_acad@abv.bg

**Abstract:** Construction of SPPs databases and SVD image compression algorithm are considered. Some examples of compression with different plates are shown.

**Keywords**: SVD, Database, Scanned astronomical photographical plates,

1  **Introduction**

　　The newly born area of Astroinformatics has emerged as an interdisciplinary area from Astronomy and modern information and communication technologies, based on the modern Internet developments. As a truly interdisciplinary area Astroinformatics has arisen from the need of information and communication technology (ICT) methods for preservation and exploitation of the scientific, cultural and historic heritage of astronomical observations. Practically, before the application of the charge-coupled devices (CCD) and orbital multi-wave telescopes the only source of information in Astronomy were the astronomical photographic observations into astronomical plates. The total information recorded on astronomical plates is estimated about 1PB (=106GB=109MB=1015Bytes) provided the information is in computer readable format. It suggests that after the plate inventory and proposed plate catalogue the plates themselves are digitized and online access to the plate image is provided. With the new development of the information and communication technologies and Internet development including the digitization and loss- lossless compression technologies of this problem can be solved completely in the near future.

　　One of the main problems is that the scanned photographic astronomical images have a huge volume (for example: the volume of one scanned astronomical plate is from 100 MB up to 2,000 MB). The access to such volume of information, as well as its analysis, is now difficult and slow task. On the other hand it is not needed for majority of tasks for analyzing the entire information available on the plate. That is why it is very essentially the digital information to be compressed in such way that the plate information to be stored and retrieved in suitable way after reduction and standardization for usage from many different users. For these purposes exceptionally useful are the methods for compression, analysis and image proceedings, developed intensively during the last 10-15 years. Their development is due mainly to the Internet progress, modern GRID technologies and GRID-based virtual portals, in particular virtual observatories.

The astronomical photographic plates (APP) obtained with a particular telescope at a particular observational site and stored at a definite place constitute the so-called plate database archive where the number ranges between several dozen to more than 100 000. Only a small number of archives have more than 10 000 plates. The largest European photographic plate database is in Sonneberg Observatory with 250 000 plates. The WIDE-FIELD PLATE DATABASE (WFPDB) is present at the Sofia Sky Archive Data Center is shown in [1] with 464,373 SPPs. The scanning process of APP includes plate images with optimal low- and high-resolution scans tabulated of Table 1 [2] and form of scanned astronomical photographic plates (SAPP) database. It is a virtual unique instrument-telescope, which allows from SAPPs searching for necessary plate index information in the available world plate vaults for the period of the last 130 years - the era of photographic astronomy.

Astronomical observations produce a large amount and a large variety of data. Many measurements of some old plates were experimented and SAPPs - however show important defects. The lifetime of archived data is nowadays a problem, since magnetic tapes have only a short lifetime (a few years), and the large amount of data involved discourages regular duplications. The SAPPs are very large data sets lead to moving to a new computer should reading data stored on the presently used media. Example, the size of some scanned images of SAPP saved in FITS files are shown in Table 2 where astronomical data usually are processing in next steps:

- First, acquired from a receptor installed on an astronomical instrument, via a command and control process; the output of the detector is called raw data (e.g. the astronomical photographic plate).

- The next process consists in detecting anomalies and removing known instrumental biases by means of calibration adjustments, and leads to calibrated or edited data.

- Last and important process is how to reduce and analyzed data expressed in physical units (e.g. fluxes as a function of wavelength). This step makes use of more sophisticated image or data processing (filtering, contrast enhancement, model fitting, etc), and frequently implies a significant reduction of the data size as compression of the scanned images.

**TABLE I**

Summarizes the number of SAPPs in Europe with flatbed scanners with high (20 microns/pixel), and low resolution;

| Observatory | Resolution | |
|---|---|---|
| | High | Low |
| Sonneberg | 215 000 | |
| Pulkovo | 40 000 | |
| Hamburg | 8 000 | |
| Vatican | 5433 | |
| Tautenburg | 4228 | |
| Asiago | 4000 | |
| Buyrakan | 2000 | |
| Potsdam | 1500 | |
| Bamberg | 1000 | 2000 |
| Bucharest | | 5000 |
| Belgrade | | 2000 |
| Sofia | | 1500 |
| Moscow-Zvenigorod | | 1000 |

**TABLE II**

Size of *.FITS files of the SAPP images with high resolution scans

| Telescope Identifier | EPSON Flatbed Scanner | Plate Size (cm.) | High Resolution (dpi) | FITS file (MB) |
|---|---|---|---|---|
| ESO100 | Expression 10000 XL | 30×30 | 2400 | 1670 |
| ROZ200 | Expression 10000 XL | 30×30 | 1600 | 664 |
| POT032 | Expression 10000 XL | 16×16 | 2400 | 440 |
| BAM10C | Perfection V750 Photo | 16×16 | 2400 | 430 |
| POT032 | Perfection V700 Photo | 16×16 | 2400 | 412 |
| BON034 | Perfection V750 Photo | 8×9 | 2400 | 126 |

## 2  Astronomical Big Data

The astronomical photographic plates constitute glass coated on one side with a dry emulsion of silver bromide. As light detectors and media for information storage, they replaced visual astronomical observations and marked the epoch of photographic astronomy that began in the 1870s and lasted for more than 130 years. Many astronomical discoveries were made on the basis of photographic plates used in the observations. Moreover, APPs stored in observatories or relevant institutions continue to be used for different astronomical tasks, especially such ones which need long time series observations as the APPs are typical examples of scientific heritage in need of preservation for future use.

Astronomical Big Data is usually defined in terms of the three V's – *Volume*, *Velocity* and *Variety* [3]:

1. *Volume*, *Velocity* – they are obvious (large data sets and data streaming or data for stars "in motion", example comets),

2. *Variety* – this is relates to the diversity of data type – this can be the considered to relate to the myriad of different objects, wavelength range/messenger type recorded, and data format as usually is used FITS format.

3. *Veracity* – this is important for analytics, and for astronomers dealing with measurements that are inherently uncertain or possibly corrupt.

The new generation of radio telescopes e.g. LOFAR [4] are often considered as good examples of Astronomical Big Data routinely produce data sets of ~ 100 TB per day, and new facilities such as the SKA [5, 6] generate object catalogues that contain billions of sources [7]. Combining this information requires a new approach to large-scale data analysis where Astronomical Big Data can also be applicable to the astronomy case, example a analytics and visualization of scanned astronomical photographical plates. Progress in data visualization has been at a virtual standstill but data analytics - new visualization approaches, automatic feature extraction tools are constructed. Many of the existing approaches to image processing are still sequential and written for single CPU cores using single threads. Now day we can use multi-threaded software for multiple CPU cores. But exist limitation - network bandwidth which is becomes a limiting factor when large volumes of data are processing and transmutation. This means that single server multi-core CPU systems may not provide enough of a performance to processing rates. The SDSS [8], which is currently in operation using a 2.5 meter telescope, reported a maximum data capture rate in the region of 200GB per day at present [9].

## 3   Present Standard Astronomical Image Formats

**FITS Formats -** The FITS format (the Flexible Image Transport System) is a standard astronomical image format endorsed by the International Astronomical Union (IAU) and NASA, which was approved in 1981. There exist libraries supported by NASA to access and manipulate the FITS file format of many languages with interfaces allowing reduction software - Python, Java, Perl, MATLAB and C++.

**IRAF -** Image Reduction and Analysis Facility is a standard application used throughout the astronomical community which has been in development since the mid to early 1980s [10], and is a Linux based software package. It provides the closest thing to a standard for data reduction and analysis within much of the astronomy community.

**NHPPS -** The NHPPS is a python based pipeline which can operate with local processing clusters of software nodes based on the OPUS system [11]. The NHPPS uses the blackboard architecture for communication across a multi-node distributed environment which is a multi-queue based system.

**ESO: Common Pipeline Library (CPL)**

CPL is Linux based operating systems which implementations of existed within various ESO instruments to astronomical data compression as require faster processors to run pipelines and does not provide multi-threaded support [12].

## 4 Computing of SVD for SAPP

## 4.1 General Theory

As an illustration we consider the SVD compression of the SAPP images. On a computer, the image is simply a matrix **A** with size $m \times n$ denoting pixel colors. Typically, such matrices can be compressed by low-rank matrices. Instead of storing the $m \times n$ entries of the matrix **A**, one need only store the $k(m+n)+k$ numbers in the sum:

$$\mathbf{A}_k = \sum_{j=1}^{k} \sigma_j \mathrm{u}_j \mathrm{v}_j^\mathrm{T}$$

where $\sigma_j$ are the singular values, $k$ - rank of the matrix, and the matrices $\mathrm{u}_j$ and $\mathrm{v}_j$ are unitary. When $k<<\min(m,n)$ this can make for a significant improvement, though modern image compression protocols use more sophisticated approaches. The SVD is stable, small perturbation in **A** correspondent to small perturbation in the matrix of singular values $\sigma_j$ and conversely. The reduced rank approximations based on the SVD are very similar in intent. However, SVD captures the best possible basis vectors for the particular data observed, rather than using one standard basis for all cases.

## 4.2 SVD for SPP database compresssion

Construction of DATABASE of compressed SAPP is show of Fig.1. First, the APPs have to scanned, and after that can be applied any image processing. Scanning process is difficult and time-consummation stage about some hours because of astronomical photographical plate images used for astronomical tasks are made at an optimal high resolution of 1600 (or 2400) dpi in FITS format, see Table 2. The scans are usually available upon request, thus the copyright of the observatories is protected. For bad quality plates only previews are needed. The systematic plate scanning takes considerable funds and gives a huge volume of scan data, which have to be stored. One possible solution of this problem is to image compression of the digitized plates.

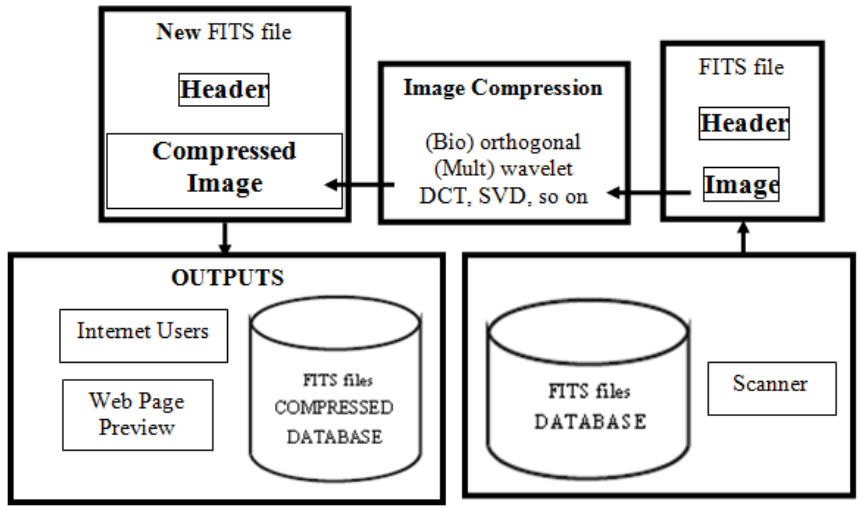

Fig.1 Construction of DATABASE of SAPP

If have already SAPP database or can download from I-net, passed to the image processing step. In this step, we applied different type spline, wavelet, curvelet, multiwavelet, DCT, SVD and so on to image compressing, or extract celestial objects, or examine of the preview image, catalog construction or other one. In [13] is propose a slight modification to the SVD, which gives us much better compression than the standard compression DCT process. For computing step we can using follow hardware resourses:

- PC with 1 procesoor
- Server with 1 procesoor
- Server with more procesoors
- Distributed Computing
- Cloud Computing

The follow step is inserting of the compression image and construct compression FITS file together header. In the output we can using compressed FITS file for Web Preview, construction or add on to existed database, or send to any I-net user.

## 5. SVD Compression of Scanned Photographical Plates

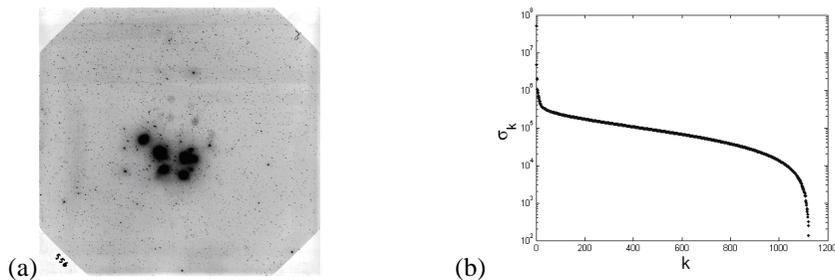

(a)                                                                 (b)

Fig.2 Image of *M45-556p.fits* in the region of the Pleiades stellar cluster a) Original image of SPP (size 1122x1122), b) Singular values

The singular values obtained from SVD for the scanned plate *M45-556p.fits* in the region of the Pleiades stellar cluster, with image size 1112x1122 pixels and 16bit format is show of Fig.2(a). We see that the singular values decrease near linearly. Though the singular values are very large, $\sigma_1 > 10^7$, there is a relative difference of five orders of magnitude between the smallest and largest singular value show in Fig.2(b). We see that the singular values decrease rapidly.

If all the singular values were roughly the same, we would not expect accurate lossless compression. We can compressed a matrix by adding only the first few terms of the series (Fig. 3).

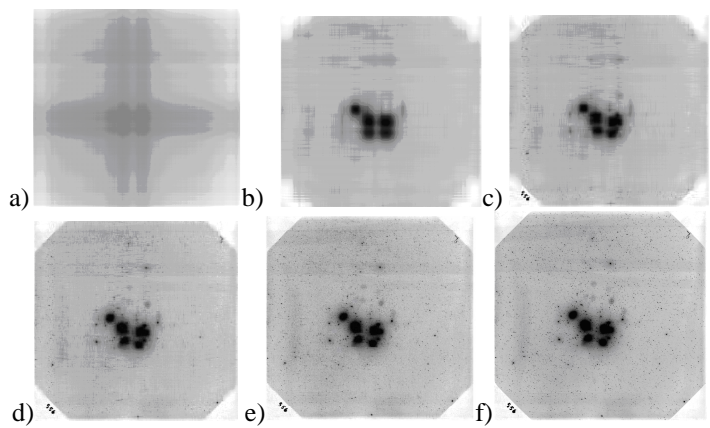

Fig.3 The SVD compression of SPP *M45-556p.fits* in the region of the Pleiades stellar cluster;
(a) k=1; (b) k=5 (c) k=12; (d) k=30; (e) k=100; (f) original image

The measure to image compressing is the compression ratio (CR) for SAPP images which is CR = Output image /Input image. Obvious, although CR>50 we recognized main ob-jects as well as show in Fig.4(a). Furthermore, for CR>100 from compressed image Fig.4(b) we easy recognized both main objects and from which SPP is this.

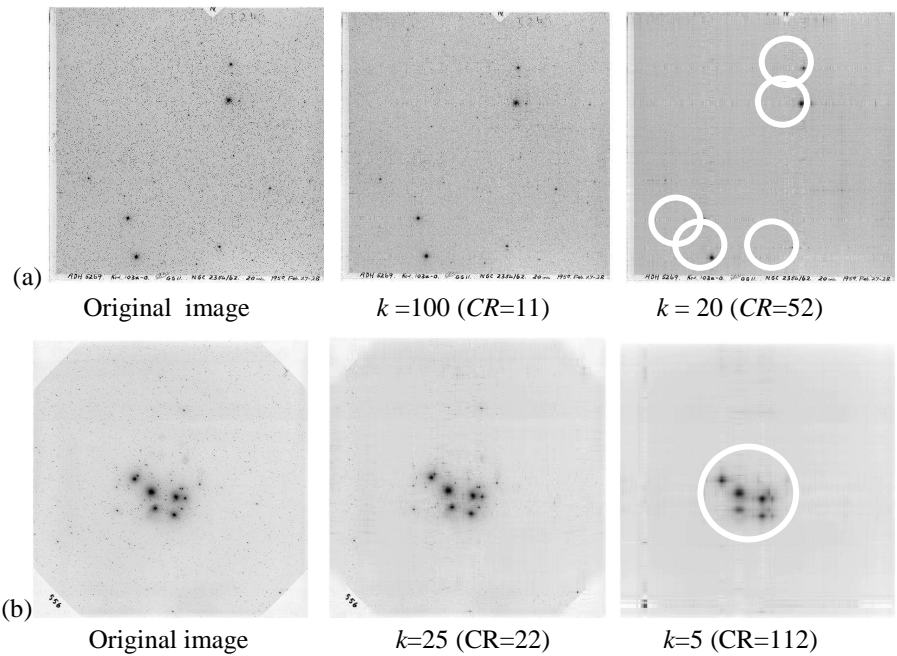

Fig.4 Compression of the image from SPP; (a) ADH5269.fits with 2048x2048 pixels; and (b) *M45-556p-1.fits* with 1120x1120 pixels.

## Conclusion

In conclusion, for Astronomical Big data great approach to decreasing volume of database is SVD compression because of for big CR the important image details from SPP are recognized. This can be use similar as filter and useful for image denoising. The SVD compression is faster from wiener filter processing. Therefore, compressed images do require less computer storage in database and transmission time than the full-rank image.